\documentstyle[preprint,eqsecnum,aps]{revtex}
\tightenlines
\begin{document}
\draft
\preprint{}
\title{
It's all in GR: spinors, time, and gauge symmetry
}
\author{A. Garrett Lisi}
\address{
Department of Physics, University of California San Diego, La Jolla, CA 
92093-0319
gar@lisi.com
}
\date{April 14, 1998}
\maketitle
\begin{abstract}
This paper shows how to obtain the spinor field and dynamics from the vielbein and geometry of General Relativity.  The spinor field is physically realized as an orthogonal transformation of the vielbein, and the spinor action enters as the requirement that the unit time form be the gradient of a scalar time field.
\end{abstract}

\newpage


\section{Introduction}

\begin{quote}
Suit the action to the word, the word to the action, with this special observance, that you o'erstep not the modesty of nature: for any thing so o'erdone is from the purpose of playing, whose end, both at the first and now, was and is, to hold as 'twere the mirror up to nature: to show virtue her feature, scorn her own image, and the very age and body of the time his form and pressure.

--HAMLET    Act 3, scene 2, 17-24
\end{quote}

The aim of this brief paper is unification, not just in the trivial sense of concatenation, but in the more general and aesthetic sense of union within an elegant framework.  A strong effort has been made towards the graceful and concise exposition of the physical and mathematical formalism, as well as towards the fluid introduction of the physical concepts.  The starting point is pseudo-Riemannian geometry, and the construction of an elegant minimal framework leads directly to the domain of Clifford algebra, a formalism allowing the seemingly facile manipulation of complex geometric entities.  A Clifford algebra decomposition of the unit basis vector frame leads naturally to the appearance of a spinor field.  A physically motivated constraint on the dynamics of the frame is suggested, that the unit time form of the frame be the gradient of a scalar time field.  This constraint produces a much sought after clock on the manifold and appears as the addition of a spinor dependent term in the gravitational action, a term corresponding to the dynamics of matter.  In this manner the unification of gravitation and matter is achieved within the most elementary framework of geometry conceivable.

\section{Manifold and Vielbein}

Assume that the universe is an $n$-dimensional pseudo-Riemannian manifold.  The geometry of the manifold is completely described in any coordinate patch by a set of orthogonal unit vectors, the vielbein, frame, or tetrad,
\begin{equation}
\hat {e_\alpha } = {(e_\alpha )}^i \vec {\partial _i}
\end{equation}
that satisfy
\begin{equation}
\hat {e_\alpha } \cdot \hat {e_\beta }  = {(e_\alpha )}^i {(e_\beta )}^j g_{ij} =  \eta _{\alpha \beta }
\label{vielb}
\end{equation}
in which $\vec {\partial _i}$ is a coordinate basis vector, roman indices are coordinate indices, $g$ is the metric, $\eta$ is the Minkowski metric, and greek indices are labels raised and lowered by $\eta$.  If a smoothly varying vielbein can be defined everywhere, the manifold is called a spin manifold.  The vielbein naturally implies a set of basis 1-forms, the fielbein
\begin{equation}
\hat{e^\alpha} = \vec{dx^i} \, ({e^{\! - \!1}}_i)^\alpha
\end{equation}
with $\vec{dx^i}$ the coordinate 1-forms, that are dual to the vielbein vectors
\begin{equation}
 {(e_\alpha )}^i {({e^{\! - \! 1}}_i )}^\beta =  \delta _\alpha^\beta
\label{fielb}
\end{equation}
or, more compactly in matrix notation, $e {e^{\! - \! 1}} = I$, where the components of $e$ are ${(e_\alpha )}^i$.  On a pseudo-Riemannian manifold the 1-forms may be identified with vectors via the metric, $\vec{dx^i} = g^{i j} \, \vec{\partial_j} = \vec{\partial^i} $ and $\hat{{e^{ - \! 1 \,}}^\alpha} = \eta^{\alpha \beta} \, \hat {e_\beta} = \hat {e^\alpha}$ , and often go under the name of covariant vectors.  The only time a distinction need be drawn between vectors and 1-forms is when taking them to a submanifold, as 1-forms are the objects that may be pulled back to a submanifold.  Any vector (the word now used interchangeably with ``form'') may be represented in terms of the coordinate or vielbein basis vectors
\begin{equation}
\vec{v} = v^i \vec{\partial _i} = v^\alpha \hat {e_\alpha } = (v^i ({e^{\! - \! 1}}_i)^\alpha) \hat {e_\alpha }
= v^i \, g_{i j} \, \vec{dx^j} = v_\alpha \hat{e^\alpha}
\end{equation}

The fielbein can be considered a factorization of the metric, since
\begin{equation}
{g}_{ij} = {({e^{\! - \! 1}}_i )}^\alpha \, \eta _{\alpha \beta } \, {({e^{\! - \! 1}}_j )}^\beta
\end{equation}
or $g = {e^{\! - \! 1}} \eta {e^{\! - \! 1}}^T$.  However, the frame of orthogonal unit vectors tangent to a manifold seems a more satisfying intuitive description than the equivalent metric, and should be interpreted as being more fundamental.  Also note that the vielbein describes an orientation on the manifold, information absent from the metric.  The metric is, however, a more compact description of the geometry, having ${n(n+1) \over 2}$ degrees of freedom compared to the vielbein's $n^2$.  The metric is invariant under local orthonormal ( Lorentz )  transformations of the fielbein,
\begin{equation}
\hat {e^\alpha} \mapsto  \hat{e^\beta} {L^\alpha}_\beta
\label{lor}
\end{equation}
with ${L^\alpha}_\beta \, \eta_{\alpha \mu} \, {L^\mu}_\nu =  \eta_{\beta \nu}$ ( or $L^T \eta L = \eta$ ), which leads to the natural unique decomposition of the fielbein matrix,
\begin{equation}
{({e^{\! - \! 1}}_i )}^\alpha =  {(\gamma_i )}^\beta {L^\alpha}_\beta
\label{fac}
\end{equation}
in which $\gamma$ is symmetric and $L$ is restricted to be a proper ( ${\rm det} \, L > 0$ ) orthochronous ( ${L^0}_0 > 0$ ) Lorentz transformation taking timelike vectors to timelike vectors and spacelike vectors to spacelike vectors, so $\hat {e^\alpha } \cdot \hat {e^\beta } = \hat {\gamma^\alpha } \cdot \hat {\gamma^\beta }$ and the direction of time is preserved.  This decomposition capitalizes on the metric invariance (\ref{lor}) to factor the fielbein into a gravitational part, the symmetric fielbein, $\hat{\gamma^\beta}$, which has ${n(n+1) \over 2}$ degrees of freedom and gives $g = {e^{\! - \! 1}} \eta {e^{\! - \! 1}}^T = \gamma \eta \gamma^T$, and the rotational part, $L$, which has ${n(n-1) \over 2}$ degrees of freedom.

\section{Spinors}

Since the 1-forms may be naturally identified with vectors via the metric, the vector dot product may be carried over this way and combined with the exterior algebra to produce a Clifford algebra with the symmetric fielbein vectors, $\hat{\gamma^\alpha}$, as 1-form basis elements satisfying the Clifford product relation,
\begin{eqnarray}
\hat{\gamma^\alpha} \hat{\gamma^\beta} & = & \hat{\gamma^\alpha} \cdot \hat{\gamma^\beta} + \hat{\gamma^\alpha} \wedge \hat{\gamma^\beta} \\
  & = & \eta^{\alpha \beta} + \hat{\gamma^\alpha} \wedge \hat{\gamma^\beta}
\end{eqnarray}
with the dot product of two vectors, $\vec{a} \cdot \vec{b} = {1 \over 2}(\vec{a} \vec{b} + \vec{b} \vec{a})$, producing a scalar and the wedge product, $\vec{a} \wedge \vec{b} = {1 \over 2}(\vec{a} \vec{b} - \vec{b} \vec{a})$, producing a 2-vector ( aka bivector or 2-form ).  Several excellent treatments of Clifford algebras and their application in physics have been made by David Hestenes and I recommend the reader to his work\cite{dh,dhgs} as an introduction.  Readers unfamiliar with Clifford algebra but familiar with Dirac matrices should note the isomorphism between the basis 1-forms and Dirac matrices, $\hat{\gamma^\alpha} \sim \gamma^\alpha$, which form a basis for the matrix representation of Clifford algebra and satisfy the same multiplicative identities, such as $\gamma^\alpha \gamma^\beta + \gamma^\beta \gamma^\alpha = 2 \eta^{\alpha \beta}$ and anti-commutivity, $\gamma^\alpha \gamma^\beta = - \gamma^\beta \gamma^\alpha$ for $\alpha \neq \beta$.

The factoring of the fielbein in (\ref{fac}) is performed in a Clifford algebra as
\begin{equation}
\hat{e^\alpha} =  \hat{\gamma^\beta} {L^\alpha}_\beta = \Psi \hat{\gamma^\alpha} \widetilde{\Psi}
\label{facv}
\end{equation}
where $\Psi$, an even or odd unitary Dirac-Hestenes spinor, is an even or odd graded multi-vector element of the Clifford algebra satisfying $\Psi \widetilde{\Psi} =1$, and $\widetilde{\Psi}$ denotes the reverse of the multi-vector, which reverses the products of all vectors in $\Psi$.  The components of $L$ may be readily obtained from $\Psi$ since
\begin{equation}
L^{\alpha \mu}  =   \hat{e^\alpha} \cdot \hat{\gamma^\mu} =  (\Psi \hat{\gamma^\alpha} \widetilde{\Psi})  \cdot  \hat{\gamma^\mu} = \hat{\gamma^\alpha} \cdot   (\widetilde{\Psi} \hat{\gamma^\mu} \Psi )
\label{rot}
\end{equation}
An even unitary spinor has ${n(n-1) \over 2}$ degrees of freedom and may be written as the exponential of a bivector, $\Psi = e^B$.

In four dimensional spacetime, $S$, an even spinor may be written out in terms of the basis as
\begin{eqnarray}
\psi & = & a_0 + b_\epsilon \hat{\gamma^0} \hat{\gamma^\epsilon} + a_\epsilon \hat{\gamma^1} \hat{\gamma^2} \hat{\gamma^3} \hat{\gamma^\epsilon} + b_0 \gamma \\
   & = & a_0 + a_\epsilon \gamma \hat{\gamma^\epsilon} \hat{\gamma^0} + \gamma (b_0 + b_\epsilon \gamma \hat{\gamma^\epsilon} \hat{\gamma^0})
\label{psi}
\end{eqnarray}
in which $\epsilon$ here sums from $1$ to $3$, and the volume element ( pseudo-scalar ) is $\gamma = \hat{\gamma^0} \hat{\gamma^1} \hat{\gamma^2} \hat{\gamma^3} = dx^0 dx^1 dx^2 dx^3 \, {\rm det } \, \gamma = {e^{\! - \! 1}} $ ( both the volume-element, $\gamma$, and symmetric fielbein matrix, $\gamma$, appear in this expression, the distinction apparent via context ).  Note that the quaternions, familiar from their use in 3-space rotations, are here equivalent to the spacelike bivectors $\gamma \sigma^\epsilon = \gamma \hat{\gamma^\epsilon} \hat{\gamma^0}$.  An even non-unitary Dirac-Hestenes spinor, $\psi$, induces a conformal transformation, a Lorentz transformation by ${L}$ and scaling by $s$ ( aka Weyl or orthogonal transformation  ), given by 
\begin{equation}
 s  \hat{\gamma^\beta} {L^\alpha}_\beta = \psi \hat{\gamma^\alpha} \widetilde{\psi}
\label{orth}
\end{equation}
in which $\psi$ is an even multi-vector free of restrictions.  An even non-unitary, non-null ( $\psi \widetilde{\psi} \neq 0$ ), spinor may be factored as $\psi = { s }^{1 \over 2} e^{\gamma {\phi \over 2}} \Psi$, which contains an even unitary spinor, $\Psi$, as well as a duality rotation, $e^{\gamma {\phi \over 2}}$, that does not effect the result of the vector transformation (\ref{orth}).  An even unitary spinor factors into a boost along $\vec{v}$ and rotation around $\vec{r}$ as $\Psi = e^B = e^{v_\epsilon \sigma^\epsilon} e^{r_\epsilon \gamma \sigma^\epsilon}$.

This is the most important fact to understand in this paper:  A unitary spinor field is defined and understood here as a Clifford algebra representation of a restricted Lorentz transformation.  And the fielbein, $\hat {e^\alpha }$, factors uniquely into a unitary spinor part, $\Psi$, and a metric part, $\hat {\gamma^\alpha }$, via (\ref{facv}).  The fielbein hence carries a spinor part and gravitational part -- spinors are already in General Relativity, they've just remained hidden in $\hat {e^\alpha }$.  This is quite different then the way spinors are usually defined, but an isomorphism holds between this definition of spinor as transformation and the standard definition.  Appendix A contains a translation to Dirac spinors and the conformal transformation matrix.  

\section{Coordinate Transformations}

Although the fielbein vectors, $\hat{e^\alpha}$, are coordinate independent objects, the fielbein matrix, ${({e^{\! - \! 1}}_i )}^\alpha$, and it's decomposition, are not coordinate independent.  A new set of $\hat{\gamma^\beta}$ and a new ${L^\alpha}_\beta$ and hence new $\Psi$ must be obtained after a coordinate transformation such that the new $\gamma$ matrix is symmetric.  This is achieved as follows:

Consider a coordinate change $x^i \rightarrow {x'}^j (x)$ that gives $\vec{dx^i} = \vec{d{x'}^j} {\partial x^i  \over \partial {x'}^j} = \vec{d{x'}^j} {L^i}_j $.  The old symmetric $\hat{\gamma^\beta}$ are now given by $\hat{\gamma^\beta} = \vec{dx^i} \, (\gamma_i)^\beta = \vec{d{x'}^j} {L^i}_j \, (\gamma_i)^\beta$.  The matrix ${L^i}_j \, (\gamma_i)^\beta$ is now not symmetric in $j$ and $\beta$.  A new set of $\hat{\gamma^\beta}'$, the old set rotated by the same $L$, is needed so that the new $\gamma'$ is symmetric.  This is
\begin{equation}
\hat{\gamma^\beta}' = \vec{d{x'}^j} \, ({\gamma'}_j)^\beta = \vec{d{x'}^j} {L^i}_j \, (\gamma_i)^\mu {L_\mu}^\beta = \hat{\gamma^\mu} {L_\mu}^\beta = \Phi \hat{\gamma^\beta} \widetilde{\Phi}
\end{equation}
giving a symmetric $\gamma' = L^T \gamma L$, and $\Phi$ the unitary spinor corresponding to transformation by $L$.  Since $\hat{e^\alpha}$ remains the same but $\hat{\gamma^\beta}$ has changed, $\Psi$ must change as well so that
\begin{equation}
\hat{e^\alpha} = \Psi \hat{\gamma^\alpha} \widetilde{\Psi} = \Psi \widetilde{\Phi} \hat{{\gamma'}^\alpha} \Phi \widetilde{\Psi} = \Psi' \hat{{\gamma'}^\alpha} \widetilde{\Psi'}
\end{equation}
and the spinor $\Psi$ changes as expected under a coordinate transformation to $\Psi' = \Psi \widetilde{\Phi}$.

\section{Derivative}

The covariant derivative, $\nabla_{\vec{V}}$, is defined to have the following properties,
\begin{eqnarray}
\nabla_{ \vec{A} + f \vec{B}} C & = & \nabla_{ \vec{A}} C+ f \nabla_{ \vec{B}} C \\
\nabla_{ \vec{A}} (B + f C) & = & \nabla_{ \vec{A}} B + f \nabla_{ \vec{A}} C + (A^i \partial_i f) C \end{eqnarray}
where $\partial_i = {\partial \over {\partial x^i}}$ and $f$ is a scalar.  The covariant derivative acting on vectors gives
\begin{eqnarray}
\nabla_i \vec{v} & = & \nabla_i ( v^k \vec{\partial_k} ) = 
( \partial_i  v^k ) \vec{\partial_k} + v^k ( \nabla_i \vec{\partial_k} ) \\
 & = & \vec{\partial_j} \ [ \partial_i  v^j + v^k g^{j m} \vec{\partial_m} \cdot ( \nabla_i \vec{\partial_k} ) ] \\
 & = & \vec{\partial_j} \ [ \partial_i  v^j + {\Gamma^j}_{ik} v^k ]
\end{eqnarray}
in which $\nabla_i = \nabla_{\vec{\partial_i}}$ and the affine connection is $\Gamma_{jik} = \vec{\partial_j} \cdot \nabla_i \vec{\partial_k}$.

Clifford algebra allows for the definition of the vector derivative, or gradient,
\begin{equation}
\vec{\nabla} = \hat{e^\alpha} \nabla_{\hat{e_\alpha}} = \hat{\gamma^\alpha} \nabla_{\hat{\gamma_\alpha}} = \hat{\gamma^\alpha} ({\gamma^{\! - \! 1}}_\alpha)^i \nabla_i = \vec{\partial^i} \nabla_i
\end{equation}
which gives, for example,
\begin{eqnarray}
\vec{\nabla} \vec{v} & = & \vec{\nabla} \cdot \vec{v}+ \vec{\nabla} \wedge \vec{v} \\
 & = & \vec{\partial^i}  \cdot \nabla_i \vec{v}+ \vec{\partial^i} \wedge \nabla_i \vec{v} \\
 & = & g^{i k}  (\partial_i v_k - {\Gamma^j}_{i k} v_j ) + \vec{\partial^i} \wedge \vec{\partial^k}  \,(\partial_i v_k - {\Gamma^j}_{i k} v_j ) \\
 & = & \hat{\gamma^\alpha} \cdot \nabla_{\hat{\gamma_\alpha}} \vec{v} + \hat{\gamma^\alpha} \wedge \nabla_{\hat{\gamma_\alpha}} \vec{v} \\
 & = & (  \partial_\mu v^\mu + {\omega^\alpha}_{\alpha \mu} v^\mu ) + \hat{\gamma^\alpha} \wedge \hat{\gamma^\beta} \,  ( \partial_\alpha v_\beta + {\omega}_{\beta \alpha \mu}  v^\mu )
\end{eqnarray}
where the spin connection for the symmetric fielbein is ${\omega}_{\beta \alpha \mu} =  \hat{\gamma_\beta} \cdot \nabla_{\hat{\gamma_\alpha}} \hat{\gamma_\mu} $ and $\partial_\alpha = {({\gamma^{\! - \! 1}}_\alpha )}^i \partial _i$ in this expression.

The covariant derivative acting on a symmetric vielbein vector may also be written as
\begin{equation}
{\nabla}_\mu \hat{\gamma_\nu} = \hat{\gamma^\alpha} \omega_{\alpha \mu \nu} = {1 \over 2} ( \omega_\mu \hat{\gamma_\nu} - \hat{\gamma_\nu} \omega_\mu )
\label{vderiv}
\end{equation}
in which the connection bivector is defined as $\omega_\mu = {1 \over 2} \omega_{\alpha \mu \nu} \, \hat{\gamma^\alpha} \wedge \hat{\gamma^\nu} $, and the shorthand ${\nabla}_\mu={\nabla}_{\hat{\gamma_\mu}}$.  This expression may be applied to an arbitrary multivector, $A$, to obtain
\begin{equation}
{\nabla}_\mu A = \bar{\partial_\mu} A + {1 \over 2} ( \omega_\mu A - A \omega_\mu )
\label{deriv}
\end{equation}
where $\bar{\partial_\mu} = ({\gamma^{\! - \! 1}}_\mu )^i \bar{\partial_i} $ is the partial derivative acting only on coefficients in $A$.  This naturally gives rise to the ``covariant spinor derivative'' when acting on objects composed of an even multiple of a spinor, for example
\begin{eqnarray}
 \nabla_{\mu} \left( \Psi \hat{\gamma^0} \widetilde{\Psi} \right)  & = &  ( \bar{\partial_\mu} \Psi + {1 \over 2} ( \omega_\mu \Psi - \Psi \omega_\mu )) \hat{\gamma^0} \widetilde{\Psi}  \\
 & + & \Psi {1 \over 2} ( \omega_\mu \hat{\gamma^0} - \hat{\gamma^0} \omega_\mu ) \widetilde{\Psi} \\
& + &  \Psi  \hat{\gamma^0} ( \bar{\partial_\mu} \widetilde{\Psi} + {1 \over 2} ( \omega_\mu \widetilde{\Psi} - \widetilde{\Psi} \omega_\mu )) \\
& = &  ( \nabla^s_{\mu} \Psi )  \hat{\gamma^0} \widetilde{\Psi} +  \Psi  \hat{\gamma^0} \widetilde{ ( \nabla^s_{\mu} \Psi )}
\end{eqnarray}
in which the middle terms cancel to give the covariant spinor derivative, $ \nabla^s_{\mu} \Psi =  \bar{\partial_\mu} \Psi + {1 \over 2} \omega_\mu \Psi$, also known as the covariant Dirac operator in curved spacetime.

\section{Curvature and Gravitation}

Since the metric is independent of local Lorentz transformations of the fielbein, all traditional geometric objects derived from the metric may be written interchangeably in terms of the fielbein, $\hat{e^\alpha}$, or symmetric fielbein, $\hat{\gamma^\alpha}$, basis.  Since the metric degrees of freedom are contained uniquely in $\hat{\gamma^\alpha}$  ( $\hat{e^\alpha}$ containing also the spin information of $\Psi$ ) traditional geometric objects will be written in terms of $\hat{\gamma^\alpha}$ except where noted.

The Ricci vectors and scalar curvature in Clifford notation are
\begin{eqnarray}
\vec{R_\alpha} & = & R_{\alpha \beta} \hat{\gamma^\beta} =   ( \vec{\nabla} \wedge \vec{\nabla} ) \cdot \hat{\gamma_\alpha}  \\
R & = & \hat{\gamma^\alpha} \cdot \vec{R_\alpha}
\end{eqnarray}
in which $R_{\alpha \beta}$ is the Ricci tensor in the symmetric fielbein basis.  Integration over the scalar curvature gives the gravitational action,
\begin{equation}
S = \int{ \gamma \ R }  = \int{ \gamma \ \left\{ \hat{\gamma^\alpha} \cdot  ( \vec{\nabla} \wedge \vec{\nabla} ) \cdot \hat{\gamma_\alpha}  \right\} }
\end{equation}
Requiring this action to be stationary with respect to independent variations of $\hat{\gamma_\alpha}$ and $\vec{\nabla}$, the Palatini method\cite{peldan}, gives the equations,
\begin{eqnarray}
\vec{R_\alpha} - {1 \over 2} \hat{\gamma_\alpha} R & = & 0 \\
 \vec{\nabla} \wedge \hat{\gamma_\alpha} +\hat{\gamma^\mu} \wedge \hat{\gamma^\nu} \, \omega_{\nu \mu \alpha} & = & 0
\end{eqnarray}
The first is the vacuum Einstein equation and the second, solvable for $\omega_{\nu \mu \alpha}$ in terms of $\hat{\gamma_\alpha}$ and $\partial_i \hat{\gamma_\alpha}$, is the defining equation for the metric compatible torsionless spin connection.

\section{Time and Spinor Dynamics}

There is a long standing problem in General Relativity, and hence in approaches to quantum gravity, regarding the nature of time.  One would like to evolve a spacelike submanifold in some coordinate time on the spacetime manifold, but the equations are invariant with respect to diffeomorphisms of the coordinates, so demanding a priori that the coordinate $x^0$ be time, a non-coordinate invariant statement, is clearly a poor option.  One needs to come up with a coordinate invariant clock, a scalar field, $t$, that, on relevant patches of the manifold, corresponds to the time.  One good option is to introduce $t$ as a separate physical scalar field on the manifold \cite{rovelli}, but consider the following alternative method:

Demand that, on relevant patches, the unit time fielbein vector, $\hat{e^0}$, be closed
\begin{equation}
\vec{\nabla} \wedge \hat{e^0} = 0
\label{closed}
\end{equation}
and hence that, on patches of the manifold for which every closed oriented curve is the boundary of some compact oriented surface\cite{frankel}, $\hat{e^0}$ is exact
\begin{equation}
\hat{e^0} = \vec{\nabla} t 
\end{equation}

This method has several good attributes.  The arrow of time, the coordinate invariant form, $\hat{e^0}$, gives rise naturally to the scalar time, $t$.  Hence the clock field, $t$, is obtained in a coordinate invariant manner using geometric elements at hand.  One may then naturally choose to transform to coordinates in which $x^0 = t$ and evolve a spacelike submanifold in Gaussian normal coordinates with $\hat{e^0}$ as the normal vector field.

The constraint equation, (\ref{closed}), is a bivector equation corresponding to the restriction of  ${n(n-1) \over 2}$ degrees of freedom and determines the dynamics of $\Psi$ as follows
\begin{eqnarray}
0 & = & \vec{\nabla} \wedge \hat{e^0} \\
 & = & \hat{\gamma^\mu} \wedge \nabla_{\mu} \left( \Psi \hat{\gamma^0} \widetilde{\Psi} \right) \\
 & = & \hat{\gamma^\mu} \wedge [ ( \nabla^s_{\mu} \Psi)  \hat{\gamma^0} \widetilde{\Psi} +  \Psi  \hat{\gamma^0} \widetilde{( \nabla^s_{\mu} \Psi )}  ] \\
 & = & 2 \hat{\gamma^\mu} \wedge < \! ( \nabla^s_{\mu} \Psi)  \hat{\gamma^0} \widetilde{\Psi}  \! >_1 \\
 & = & 2 \hat{\gamma^\mu} \wedge \hat{\gamma^\nu} < \! \hat{\gamma_\nu} ( \nabla^s_{\mu} \Psi)  \hat{\gamma^0} \widetilde{\Psi} \! >_0 \\
 & = & 2 \hat{\gamma^\mu} \wedge \hat{\gamma^\nu} < \! \overline{\Psi} \hat{\gamma_\nu} ( \nabla^s_{\mu} \Psi) \!  >_0 \\
 & = & 2 \hat{\gamma^\mu} \wedge \hat{\gamma^\nu} \, T_{\mu \nu}
\end{eqnarray}
in which the operator $<>_n$ gives the grade $n$ elements of a multivector, $\overline{\Psi} = \hat{\gamma^0} \widetilde{\Psi}$, and the energy-momentum tensor for a spinor field, $T_{\mu \nu}$, has been recognized.  The requirement that the unit time vector, $\hat{e^0}$, be closed is hence equivalent to the requirement that the anti-symmetric part of the spinor energy-momentum tensor, $T_{\mu \nu}$, vanish.

One natural way to enforce the vanishing of $T_{[\mu \nu]}$ is to construct the equations of motion to be
\begin{equation}
R_{\mu \nu} - {1 \over 2} \eta_{\mu \nu} R  =  T_{\mu \nu}
\end{equation}
in which the symmetry of the Ricci tensor will enforce the vanishing of $T_{[\mu \nu]}$.  Since $T_{\mu \nu} = \, < \! \overline{\Psi} \hat{\gamma_\nu} {({\gamma^{\! - \! 1}}_\mu )}^i ( \nabla^s_i \Psi) \! >_0 $ is the energy-momentum tensor of the standard spinor action, these equations of motion will come from the action
\begin{equation}
S = \int{ \gamma \ \left\{ R + < \! \overline{\Psi} \vec{\nabla^s} \Psi \!  >_0  \right\} }
\label{gmac}
\end{equation}
Thus it appears that the matter action arises from the geometric restriction that full fielbein gravity have a closed unit time vector.  To vary $\Psi$ in (\ref{gmac}) in spacetime one may vary $\Psi$ over all even multi-vectors and enforce $\Psi \widetilde{\Psi} =1$ via the method of Lagrange multipliers.  Introducing $m_s$ and $m_p$ as Lagrange multiplier scalar fields, (\ref{gmac}) is equivalent to
\begin{equation}
S = \int{ \gamma \ \left\{ R + < \! \overline{\psi} \vec{\nabla^s} \psi \!  >_0 + < \! (m_s + \gamma m_p  )(\psi \widetilde{\psi} - 1)  \! >_0  \right\} }
\label{mgmac}
\end{equation}
with $\psi$ varied over it's eight degrees of freedom and the physical interpretation of $m$ clear.  If $\psi$ is dynamically restricted to be unitary, as above, then one might also consider terms in the action such as $\psi R \widetilde{\psi}$ and address conformal  symmetry. 

An alternative to the inclusion of the term $< \! \overline{\Psi} \vec{\nabla^s} \Psi \!  >_0$ in the action might be the direct restriction to $\vec{\nabla} \wedge \hat{e^0} = 0$ via the method of Lagrange multipliers, with the inclusion of a term such as $< \! B_2 \vec{\nabla} ( \Psi \hat{\gamma^0} \widetilde{\Psi} )  \!  >_0$, with $B_2$ a Lagrange multiplier bivector field and dynamical selection of $B_2$ acting as a method of symmetry breaking.

\section{Gauge Symmetries}

A central proposal of this paper is that the ostensibly non-gravitational dynamics of the fielbein, the dynamics of $\psi$, are contained entirely in the geometric restriction $\vec{\nabla} \wedge \hat{e^0} = 0$.  The dynamics imposed by (\ref{mgmac}) on $\psi$ are overly restrictive in this regard and need to be loosened by the addition of symmetries corresponding to the symmetries of $\vec{\nabla} \wedge \hat{e^0} = 0$.  This is accomplished via the method of adding vector gauge fields and couplings to attain the necessary symmetries.

The first symmetries to notice in $\vec{\nabla} \wedge \hat{e^0} = \hat{\gamma^\mu} \wedge \nabla_{\mu} \left( \psi \hat{\gamma^0} \widetilde{\psi} \right) = 0$ are the invariance of this equation under duality and space rotations of $\hat{\gamma^0}$.  The duality rotation invariance is invariance of  $\hat{e^0}$ under the transformation $\psi \mapsto \psi e^{\gamma {\phi \over 2}}$.  It corresponds to the group $U(1)$ and necessitates the addition of the corresponding vector gauge field in the standard manner to (\ref{mgmac}).  The space rotation invariance is invariance of  $\hat{e^0}$ under the transformations $\psi \mapsto \psi e^{r_\epsilon \gamma \sigma^\epsilon}$.  It corresponds to the group $SU(2)$ and necessitates the addition of the three corresponding gauge fields.

One should consider the lifting of the restriction that $\psi$ in (\ref{mgmac}) be even.  This raises the possibility that one might add to $\psi$ an odd multivector part, such as $\psi_{\! o} \, e_L$, in which $\psi_{\! o}$ is odd and $e_L = {1 \over 2} ( 1 + \hat{\gamma^3} \hat{\gamma^0} )$ is an idempotent projection operator.  This ensures that the scalar and pseudo-scalar parts of $\psi \widetilde{\psi}$ remain unchanged, since $e_L \widetilde{e_L} = 0$.  Note also that $e_L \hat{\gamma^0} \widetilde{e_L}$ is a null vector, so factoring $\psi$ into the left ideals of $e_L$ and $\widetilde{e_L} = e_R$ shows how $\psi$ is built from left and right chirality states and allows one to see the geometric meaning of $U(1) \times SU(2)_L$.

A possible symmetry of $\vec{\nabla} \wedge \hat{e^0} = 0$ to notice is the symmetry of transformations of the whole bivector equation, $B = \vec{\nabla} \wedge \hat{e^0} = 0$.  This equation will be satisfied if and only if $< \! B \widetilde{B} \! >_0 \, = 0$.  The scalar  $< \! B \widetilde{B} \! >_0$ is invariant under the group $SU(3)$ of transformations of the bivector $B$, as described in \cite{hest3}, though it does not seem possible to incorporate this symmetry into $\psi$ as a gauge symmetry in the same manner as $U(1)$ and $SU(2)$.  This difficulty suggests that it may be necessary to consider other dimensions from the four of spacetime in order to naturally obtain the $U(1) \times SU(2) \times SU(3)$ gauge symmetry of the standard model.

\section{Conclusion}

In this paper a step towards unification has been achieved, the unification of matter and gravity in a minimal geometric framework.  Although others have recently proposed a similar unification scheme of obtaining gravitational dynamics from the Dirac operator \cite{connes}, the path proposed in the present exhibition goes in the other direction by obtaining the spinor field and Dirac operator from geometry.

A significant problem remaining in the current approach is the difficulty in satisfactorily accommodating $SU(3)$ symmetry.  It seems plausible that $SU(3)$ could obtain with the addition of other dimensions to spacetime, perhaps in a Kaluza-Klein compactification scheme or in a holographic approach.  And, of course, this program on the unification of matter and gravity would be incomplete without mention of the possibility of understanding quantum field theory within a geometric model.  It is my greatest hope that this goal will be achieved, and that this work has furthered progress towards that end.

\newpage 

\appendix
\section{Spinors and Orthogonal Transformations}

An even Dirac-Hestenes spinor, $\psi$, in spacetime may be written as a $4\times 4$ matrix if the $\hat{\gamma^\alpha}$ are identified with Dirac matrices and multiplied accordingly.  For a Dirac matrix choice, in agreement with a mostly negative $\eta$, of
\begin{eqnarray*}
{\gamma}^{0} = \pmatrix{
I & 0 \cr
0 & -I \cr
} \quad
{\gamma}^\epsilon =  \pmatrix{
0 & {\sigma}^\epsilon \cr
-{\sigma}^\epsilon & 0 \cr
}
\end{eqnarray*}
in which $I$ is the 2$\times$2 identity matrix and ${\sigma}^\epsilon$ 
represents the Pauli matrices
\begin{eqnarray*}
{\sigma}^{1} = \pmatrix{
0 & 1 \cr
1 & 0 \cr
} \quad
{\sigma}^{2} = \pmatrix{
0 & -i \cr
i & 0 \cr
} \quad
{\sigma}^{3} = \pmatrix{
1 & 0 \cr
0 & -1 \cr
}
\end{eqnarray*}
(\ref{psi}) gives
\begin{eqnarray}
\psi = \pmatrix{
a_0 + i \ a_3 & a_2 + i \ a_1 & b_3 - i \ b_0 & b_1 - i \ b_2 \cr
-a_2 + i \ a_1 & a_0 - i \ a_3 & b_1 + i \ b_2 & -b_3 - i \ b_0 \cr
b_3 - i \ b_0 & b_1 - i \ b_2 & a_0 + i \ a_3 & a_2 + i \ a_1 \cr
b_1 + i \ b_2 & -b_3 - i \ b_0 & -a_2 + i \ a_1 & a_0 - i \ a_3 \cr
}
\end{eqnarray}
Each column and each row of the matrix spinor representation contains all of the information in $\psi$.  The translation can be made to the equivalent Dirac column spinor by
\begin{eqnarray}
\psi_D = \psi \pmatrix{
1 \cr
0 \cr
0 \cr
0 \cr
} =
 \pmatrix{
a_0 + i \ a_3 \cr
-a_2 + i \ a_1\cr
b_3 - i \ b_0 \cr
b_1 + i \ b_2 \cr
}
\end{eqnarray}
which contains the same information as the spinor $\psi$ in (\ref{psi}), though in a less intuitive representation.  Since the massless Dirac equation is equivalent in either notation, $\hat{\gamma^\beta} \bar{\partial_\beta} \psi = 0 \Leftrightarrow \gamma^\beta \partial_\beta \psi_D = 0$,  translation from one notation to the other is immediate.  Note also the relationship $\widetilde{\psi} = \gamma_0 \overline{\psi} = \gamma_0 \psi^\dagger \gamma^0 $.

The matrix elements of $s L^{\alpha \beta}$ can be calculated via (\ref{rot},\ref{orth}) in terms of the coefficients of $\psi$, and hence in terms of the Dirac spinor, $\psi_D$.  They are presented here to convince the reader of this identification.
\begin{eqnarray*}
s L^{0 0} & = & {a_0}^2+{a_3}^2+{a_2}^2+{a_1}^2+{b_3}^2+{b_0}^2+{b_1}^2+{b_2}^2 \\
s L^{0 1} & = & 2(- {a_0} {b_1}+ {a_3} {b_2}+ {a_1} {b_0}- {a_2} {b_3}) \\
s L^{0 2} & = & 2(- {a_3} {b_1}- {a_0} {b_2}+ {a_2} {b_0}+ {a_1} {b_3}) \\
s L^{0 3} & = & 2( {a_2} {b_1}- {a_1} {b_2}+ {a_3} {b_0}- {a_0} {b_3}) \\
s L^{1 0} & = & 2( {a_0} {b_1}+ {a_3} {b_2}- {a_2} {b_3}- {a_1} {b_0}) \\
s L^{1 1} & = & -{a_0}^2+{a_3}^2-{a_1}^2+{a_2}^2-{b_0}^2+{b_3}^2-{b_1}^2+{b_2}^2 \\
s L^{1 2} & = & 2(- {b_0} {b_3}- {b_1} {b_2}- {a_0} {a_3}- {a_1} {a_2}) \\
s L^{1 3} & = & 2( {b_0} {b_2}- {b_3} {b_1}+ {a_0} {a_2}- {a_3} {a_1}) \\
s L^{2 0} & = & 2(- {a_3} {b_1}+ {a_0} {b_2}+ {a_1} {b_3}- {a_2} {b_0}) \\
s L^{2 1} & = & 2( {a_0} {a_3}- {a_1} {a_2}+ {b_0} {b_3}- {b_1} {b_2}) \\
s L^{2 2} & = & -{a_0}^2+{a_3}^2+{a_1}^2-{a_2}^2-{b_0}^2+{b_3}^2+{b_1}^2-{b_2}^2 \\
s L^{2 3} & = & 2(- {b_0} {b_1}- {b_3} {b_2}- {a_0} {a_1}- {a_3} {a_2}) \\
s L^{3 0} & = & 2(- {a_3} {b_0}+ {a_0} {b_3}+ {a_2} {b_1}- {a_1} {b_2}) \\
s L^{3 1} & = & 2(- {a_3} {a_1}- {a_0} {a_2}- {b_3} {b_1}- {b_0} {b_2}) \\
s L^{3 2} & = & 2( {a_0} {a_1}- {a_3} {a_2}+ {b_0} {b_1}- {b_3} {b_2}) \\
s L^{3 3} & = & {b_1}^2+{b_2}^2-{b_0}^2-{b_3}^2+{a_1}^2+{a_2}^2-{a_0}^2-{a_3}^2
\end{eqnarray*}
A symbolic computation confirms that this satisfies  $s L^T \eta s L = s^2 \, \eta$ and gives the scaling
\begin{eqnarray*}
s^2 = s^2 L^T \eta L \eta = {a_0}^4+{a_1}^4+{a_2}^4+{a_3}^4+{b_0}^4+{b_1}^4+{b_2}^4+{b_3}^4 \\
+2 {a_0}^2 {b_0}^2-2 {a_0}^2 {b_1}^2 -2 {a_0}^2 {b_2}^2+2 {a_3}^2 {a_2}^2+2 {a_3}^2 {a_1}^2+2 {a_3}^2 {b_3}^2 \\
-2 {a_3}^2 {b_0}^2-2 {a_3}^2 {b_1}^2-2 {a_3}^2 {b_2}^2+2 {a_2}^2 {a_1}^2-2 {a_2}^2 {b_3}^2-2 {a_2}^2 {b_0}^2 \\
-2 {a_2}^2 {b_1}^2+2 {a_2}^2 {b_2}^2-2 {a_1}^2 {b_3}^2-2 {a_1}^2 {b_0}^2+2 {a_1}^2 {b_1}^2-2 {a_1}^2 {b_2}^2 \\
+2 {a_0}^2 {a_1}^2-2 {a_0}^2 {b_3}^2+2 {b_3}^2 {b_0}^2+2 {b_0}^2 {b_2}^2+2 {b_0}^2 {b_1}^2
+2 {b_3}^2 {b_2}^2 \\
+2 {a_0}^2 {a_3}^2+2 {a_0}^2 {a_2}^2+2 {b_3}^2 {b_1}^2+2 {b_1}^2 {b_2}^2+8 {a_0} {b_1} {a_1} {b_0} \\
+8 {a_0} {b_2} {a_2} {b_0}+8 {a_2} {b_1} {a_1} {b_2}+8 {a_3} {b_0} {a_0} {b_3}+8 {a_3} {b_1} {a_1} {b_3}+8 {a_3} {b_2} {a_2} {b_3}
\end{eqnarray*}

\narrowtext
\end{document}